\documentclass[a4paper]{article}

\usepackage{INTERSPEECH2018}

\usepackage{anyfontsize}
\usepackage{t1enc}

\usepackage{graphicx}
\usepackage{amssymb,amsmath,bm}
\usepackage{textcomp}
\usepackage{color}
\usepackage{multirow}
\usepackage{hyperref}
\usepackage{url}
\usepackage{subcaption}
\usepackage{soul}
\usepackage{cleveref}
\usepackage{graphicx}

\usepackage{soul}  

\def\vec#1{\ensuremath{\bm{{#1}}}}

\PassOptionsToPackage{dvipsnames,table}{xcolor}
\usepackage{pgfplotstable}
\pgfplotsset{compat=1.14}

\ninept

\title{COVID-19 and Computer Audition: An Overview on What \\ Speech \& Sound Analysis Could Contribute in the SARS-CoV-2 Corona Crisis}

\makeatletter
\def\name#1{\gdef\@name{#1\\}}
\makeatother \name{\textit{Bj\"orn W.\ Schuller$^{1,2,3}$}, Dagmar M.\ Schuller$^{3}$, Kun Qian$^{4}$, Juan Liu$^{5}$, Huaiyuan Zheng$^{6}$, Xiao Li$^{7}$}

\address{\fontsize{11}{11}\selectfont 
    $^1$GLAM -- Group on Language, Audio \& Music, Imperial College London, UK \\
    $^2$EIHW -- Chair of Embedded Intelligence for Health Care and Wellbeing, University of Augsburg, Germany \\
	$^3$audEERING GmbH, Gilching, Germany \\
	$^4$Educational Physiology Laboratory, The University of Tokyo, Tokyo 113-0033, Japan \\
	$^5$Department of Plastic Surgery, The Central Hospital of Wuhan, Tongji Medical College,\\Huazhong University of Science and Technology, Wuhan 430014, P.\,R.\,China \\
	$^6$Department of Hand Surgery, Wuhan Union Hospital, Tongji Medical College,\\Huazhong University of Science and Technology, Wuhan 430022, P.\,R.\,China \\
	$^7$Department of Neurology, Children's Hospital of Chongqing Medical University,\\Chongqing Medical University, Chongqing 400014, P.\,R.\,China
}

\email{schuller@IEEE.org}

\begin{document}

\maketitle
\begin{abstract}
At the time of writing, the world population is suffering from more than 10\,000 registered COVID-19 disease epidemic induced deaths since the outbreak of the Corona virus more than three months ago now officially known as SARS-CoV-2. Since, tremendous efforts have been made worldwide to counter-steer and control the epidemic by now labelled as pandemic. In this contribution, we provide an overview on the potential for computer audition (CA), i.\,e., the usage of speech and sound analysis by artificial intelligence to help in this scenario. We first survey which types of related or contextually significant phenomena can be automatically assessed from speech or sound. These include the automatic recognition and monitoring of breathing, dry and wet coughing or sneezing sounds, speech under cold, eating behaviour, sleepiness, or pain to name but a few. Then, we consider potential use-cases for exploitation. These include risk assessment and diagnosis based on symptom histograms and their development over time, as well as monitoring of spread, social distancing and its effects, treatment and recovery, and patient wellbeing. We quickly guide further through challenges that need to be faced for real-life usage. We come to the conclusion that CA appears ready for implementation of (pre-)diagnosis and monitoring tools, and more generally  provides rich and significant, yet so far untapped potential in the fight against COVID-19 spread.
\end{abstract}


\section{Introduction}
\label{Introduction}

The World Health Organisation's (WHO) office in China was first made aware of the previously unknown SARS-CoV-2 `Corona' virus on the last day(s) of the last year, i.\,e., roughly three and a half month ago. On 11 March 2020, the WHO declared the disease triggered by the virus -- COVID-19 -- as pandemic. The spread of the disease induced by the SARS-CoV-2 or `Corona' virus is assumed to underlie an exponential growth.
Johns Hopkins University's Center for Systems Science and Engineering (CSSE) gathers via several sources the reported official Corona virus COVID-19 Global Cases\footnote{assessed on 20 March 2020, 11:13:39 AM server time} which at the time of writing were assumed to account for 245\,484 with 10\,031 deaths, and 86\,035 reported as recovered. However, whether there are long-term effects after recovery is yet to be researched. In the light of this dramatic spread, one is currently internationally witnessing drastic counter-measures that have not been seen in this form over decades in many countries. These include significant public `shut-down' measures to foster `social distancing' in order to slow-down and control further spread.

As research globally is making massive efforts to contribute to better understand and fight the phenomenon from a medical and interdisciplinary point of view, also computer science and engineering in terms of `Digital Health' solutions aim at maximum exploitation of available and realisable means. In particular, in combination with Artificial Intelligence (AI), one can exploit a powerful tool, which so far has largely been tapped for prediction of COVID-19 spread (cf.\, e.\,g., \cite{hu2020artificial}) and Computer Vision (CV) approaches in the Corona context such as for automatic screening for COVID-19 on CT images \cite{gozes2020rapid,wang2020deep}. There is, however, broader belief that also other signals including such from sensors on a smartphone could help even in the diagnosis of COVID-19  \cite{maghdid2020novel}.

In the following, we aim to provide an overview on what Computer Audition (CA), i.\,e., the application of computing for audio processing including `Machine Listening', `Computational Paralinguistics', and more general speech and sound analysis, but also synthesis, could contribute in this situation. To the best of the authors' knowledge, this resource is so far not used despite offering a plethora of opportunities in this context. 

The remainder of this overview is structured as follows: We first summarise phenomena more and less closely related to the case of COVID-19 that have already been targeted by CA and would be readily available. Examples include automatic recognition of speakers suffering from a cold or wearing a mask, breathing, coughing and sneezing sounds, or recognition of audio in spatial proximity. We then shift to the introduction of concrete use-cases how CA could benefit the ongoing global fight against the Corona crisis. Subsequently, we introduce challenges and entry barriers from a technical as well as ethical and societal point of view before concluding this overview.

\section{Computer Audition: Related Phenomena}
In the following, we set out by show-casing what CA has already successfully targeted as audio cases for recognition, and appears related to the task of interest in this contribution -- fighting the ongoing COVID-19 spread. 

\subsection{Speech Analysis}

Speech analysis by computational means is highly related to the field of Computational Paralinguistics \cite{Schuller13-CPE}. The field has several related recognition tasks on offer. These are often well documented in the framework of competitive challenge events such as the Interspeech Computational Paralinguistics Challenge (ComParE). The latter has -- perhaps closest related to the COVID-19 case -- in its 2017 edition featured the automatic recognition of speech under cold \cite{Schuller17-TI2}, i.\,.e., automatically recognising speakers affected by a cold from the acoustics of their voice. In the ongoing challenge of this year, the continuous assessment of breathing patterns from the speech signal appears relevant \cite{Schuller20-TI2}, e.\,g., as basis to recognise COVID-19's often witnessed symptoms of short-breathiness and breathing difficulties. The current ComParE challenge further targets the recognition of speech under mask, i.\,e., the automatic recognition whether a speaker is wearing a facial protective mask, and the recognition of emotion of elderly, which may become interesting in monitoring the aftermath of social isolation of elderly, as currently discussed, e.\.g., in the U.\,K. for three months. On the age scale's opposite end, toddlers' crying sounds seems to be the only indicator to understand if they are suffering from COVID-19 symptoms. In the ComParE challenge series, infant crying was investigated in 2018 \cite{Schuller18-TI2}, and the valence, i.\,e., positivity of baby sounds in 2019 \cite{Schuller19-TI2}. As symptoms of COVID-19 can also include lack of appetite, it seems further interesting to reference to the EAT challenge \cite{Schuller20-TCO}: In this event, it was demonstrated that one can infer from audio whether speech under eating and eating sounds indicate eating difficulty and ``likability'' related to whether one enjoys eating. The assessment of sleepiness -- a further symptom of COVID-19 -- was first featured in ComParE in 2011 \cite{Schuller14-MSS} as binary task, and  as continuous sleepiness assessment on the Karolinska sleepiness scale in 2019 \cite{Schuller19-TI2}. Also pain such as headache or bodily pain can accompany COVID-19; speech under pain has also been shown to be automatically accessibly \cite{oshrat2016speech,Ren18-EOT}. When it comes to individual risk assessment and monitoring, speaker traits may be of interest. High mortality risk groups include the elderly, and (slightly) higher mortality rate was so far seen in male individuals \cite{caramelo2020estimation}. Age and gender were also shown in the context of ComParE, and can be automatically determined reliably given sufficient speech material \cite{Weninger12-IRO}. A history of health issue can further indicate high risk. A number of health-related speaker state information relevant in this context has been shown feasible such as individuals suffering from asthma \cite{mazic2015two}, head-and-neck cancer \cite{maier2009automatic}, or smoking habits \cite{poorjam2014multitask,satori2017voice}.

Speaker diarisation, i.\,e., determining who is speaking when, and speaker counting \cite{xu2013crowd++} can become of interest in the ongoing social distancing. When it comes to counter measures such as quarantine, or risk assessment of individuals, one could also consider the usage of automatic recognition of deceptive speech when people are questioned about their recent contacts or whereabouts, as their personal work and life interests may interfere with the perspective of being sent to quarantine.  Deception and sincerity were targeted in ComParE in 2016 \cite{Schuller16-TI2}.
Monitoring wellbeing of individuals during social distancing and quarantine can further find interest in depression and fear recognition. Both were shown feasibly to be assessed from speech in the Audio/Visual Emotion Challenge (AVEC) challenge event series \cite{Valstar16-SFA} including from speech only at reasonable deviation on a continuous scale.

Generally speaking, speech audio also includes textual cues. Broadening up to Spoken Language Processing (SLP), this can also be of help to gather and analyse information from spoken conversations available in individual communications, news or social media. For textual cues, this has already been considered \cite{pandey2020machine}. From a speech analysis perspective, this includes Automatic Speech Recognition (ASR) and Natural Language Processing (NLP).

\begin{table}[t!]
  \caption{Interdependence of Computer Audition (CA) tasks and potential use-cases in the context of the Corona crisis. SLP: Spoken Language Processing. Health state encompasses a wider range of health factors such as asthma, head and neck cancer, or smoking habits that can be inferred from speech audio.}
  \label{tab:task-ucase}
  \centering
  \begin{tabular}{ l | r r r}
    \toprule
    CA Task & \rotatebox{90}{Risk Assessment} & \rotatebox{90}{Diagnosis} & \rotatebox{90}{Monitoring} \\
    \midrule
    \multicolumn{4}{c}{\textbf{Speech Analysis}} \\
    \midrule
    Age \& Gender                   & $\surd$   &           &           \\
    Breathing                       &           & $\surd$   & $\surd$   \\
    Cold                            &           & $\surd$   &           \\
    Crying (Infants)                &           & $\surd$   &           \\
    Deception \& Sincerity          &           &           & $\surd$   \\   
    Depression                      &           &           & $\surd$   \\   
    Emotion (incl.\ of Elderly)     &           &           & $\surd$   \\   
    Health State                    & $\surd$   &           & $\surd$   \\   
    Lung Sounds                     &           & $\surd$   & $\surd$   \\   
    Mask                            & $\surd$   &           & $\surd$   \\   
    Pain                            &           & $\surd$   & $\surd$   \\   
    Sleepiness                      &           & $\surd$   & $\surd$   \\    
    SLP                             &           &           & $\surd$   \\   
    Speaker Count                   & $\surd$   &           & $\surd$   \\   
    \midrule
    \multicolumn{4}{c}{\textbf{Sound Analysis}}     \\
    \midrule
    Coughing (dry, wet, productive) &           & $\surd$   & $\surd$   \\   
    Cardiovascular Disease          & $\surd$   &           & $\surd$   \\   
    Diarisation                     & $\surd$   &           & $\surd$   \\   
    Localisation                    & $\surd$   &           & $\surd$   \\   
    Proximity                       & $\surd$   &           & $\surd$   \\   
    Sneezing                        &           & $\surd$   & $\surd$   \\   
    Snoring                         &           & $\surd$   & $\surd$   \\   
    Source Separation               & $\surd$   &           & $\surd$   \\     
    Swallowing                      &           & $\surd$   & $\surd$   \\   
    Throat Clearing                 &           & $\surd$   & $\surd$   \\   
    \bottomrule
  \end{tabular}
\end{table}

\subsection{Sound Analysis}
From a sound analysis perspective, one may first consider such of interest for COVID-19 use-cases that are produced by the human body. In the context of COVID-19, this includes fore mostly the automatic recognition of coughs \cite{matos2006detection,olubanjo2014tracheal,Amiriparian17-CAD} including dry vs wet coughing \cite{moradshahi2012improving} and dry vs productive coughing \cite{schroder2016classification} and sneeze \cite{Amiriparian17-CAD}, swallowing and throat clearing \cite{olubanjo2014tracheal} sounds -- all showcased at high recognition rates.
As severe COVID-19 symptoms are mostly linked to developing a pneumonia which is the cause of most deaths of COVID-19, it further appears of interest that different breathing patterns, respiratory sounds and lung sounds of patients with pneumonia can be observed through CA \cite{murphy2004automated}, even with mass devices such as smart-phones \cite{song2015diagnosis}. 
Of potential relevance could also be the already possible monitoring of different types of snoring sounds \cite{Janott18-SCT} including their excitation pattern in the vocal tract and their potential change over time to gain insight on symptoms also during sleep.
Further, highest risk of mortality from COVID-19 has been seen for such suffering from cardiovascular disease followed by chronic respiratory disease. In ComParE 2018, heart beats were successfully targeted from audio for three types heart status, namely, normal, mild, and moderate/severe abnormality.
Hearing local proximity from ambient audio further appears possible \cite{Pokorny19-SAT}, and could be used to monitor individuals potentially too close to each other in the `social distancing' protective counter-measure scenarios.
3D audio localisation \cite{delikaris20163d} and diarisation further allows for locating previously recognised sounds and attributing them to to sources. This could further help in the monitoring of public spaces or providing warnings to users as related to individuals potentially being locally too close with directional pointers.
Audio source separation and denoising \cite{Liu19-NIT} of stethoscope sounds and audio \cite{yang2013heart} for clinicians and further processing can additionally serve as tool.


\section{Potential Use-Cases}
Let us next elaborate on use-cases we envision as promising for CA in the context of COVID-19.
A coarse visual overview on the dependence of CA tasks and these use-cases is provided in Table \ref{tab:task-ucase}. Check-marks indicate that the already available automatic audio analysis tasks listed in the left column appear of interest in the three major use-case groups listed in the right-most three columns. Note that these are indicative in nature.

\subsection{Risk Assessment}
A first use-case targets the prevention of COVID-19 spread by individual risk assessment. As shown above, speaker traits like age, gender, or health state can be assessed automatically from the voice to provide an estimate on the individual mortality risk level. In addition, one can monitor if oneself or others around are wearing a mask when speaking, count speakers around oneself and locate these and their distance to provide a real-time ambient risk assessment and informative warning.

\subsection{Diagnosis}
While the standard for diagnosis of COVID-19 is currently a nasopharyngeal swab, several other possibilities exist including chest CT-based analysis as very reliable resource as outlined above. Here, we consider whether an audio-based diagnosis could be possible. While it seems clear that such an analysis will not be suited to compete with the state-of-the-art in professional testing previously named, its non-invasive and ubiquitously available nature would allow for individual pre-screening `anywhere', `anytime', in real-time, and available more or less to `anyone'. To the best of the authors' knowledge, no study has yet investigated audio from COVID-19 patients vs highly varied control group data including such suffering from influenza or cold and healthy individuals. Unfortunately, coughing and sneezing of COVID-19 patients does not differ significantly to human perception from `normal' patients. This includes lung and breathing sounds.
However, \cite{wang2020abnormal} assume that abnormal respiratory patterns can be a clue for diagnosis.
Overall, by that, it seems unclear if diagnosis from short audio samples of patients could be directly possible, given that most speech or body sounds are likely not to show significant differences for closely related phenomena such as influenza or cold. 

Rather, we believe that a histogram of symptoms over time in combination with their onset appears highly promising. Table \ref{tab:histrogram} visualises this concept in a qualitative manner by coarse ternary quantification of each symptom or `feature' from a machine learning perspective\footnote{based on  https://www.qld.gov.au, https://www.medicinenet.com/, https://www.medicalnewstoday.com, all assessed on 20 March 2020.}. Each of the symptoms in the table can -- as outlined above -- (already) be assessed automatically from an intelligent audio sensor. In a suited personal application such as on a smart phone or smart watch, smart home device with audio abilities, or via a telephone service, etc., one could collect frequency of symptoms over time and from the resulting histogram differentiate with presumably high success rate between COVID-19, influenza, and cold. By suited means of AI, a probability could be given to users how likely their symptoms speak for COVID-19. Of particular interest thereby is also the ``Onset Gradient'' feature in Table \ref{tab:histrogram}. It alludes to whether the onset of symptoms over time is gradual (i.\,e., over the span of up to two weeks or more) or rather abrupt (i.\,e., within hours or a few days only), which can be well observed by AI analysis in a histogram sequence updated over time. Collecting such information from many users, this estimate for histogram-based diagnosis of COVID-19 can be improved in precision over time if users ``donate their data''. In addition, clinicians could be given access to the histogram or be pointed to typical audio examples in a targeted manner remotely that have been collected over longer time to speed up the decision whether the users should go for other more reliable forms of testing. This could help to highly efficiently pre-select individuals for screening.

\begin{table}[t!]
  \caption{Qualitative behaviour of symptoms of COVID-19 vs cold and influenza (flu): Tentative histogram by symptom (`feature'/`variable') in ternary quantification (from no/low (`+') to frequent/high (``+++')). Shown is also the symptom gradient onset behaviour. Further frequently related variables include diarrhea, fever, or watery eyes, which could partially be assessed also by audio -- the latter two rater by physiological and visual sensors, respectively. Assembled from diverse references. The table is indicative in nature on purpose, and more fine-grained quantification could apply.}
  \label{tab:histrogram}
  \centering
  \begin{tabular}{ l | r r r}
    \toprule
    Symptom & \rotatebox{90}{COVID-19} & \rotatebox{90}{Influenza} & \rotatebox{90}{Cold} \\
    \midrule
Breathing: Dypnea (Shortness)	    &+++        &++         &+  \\
Breathing: Difficulty	            &+++        &++	        &+  \\
    \midrule
Rhinorrhea (Runny Nose)	            &+	        &++	        &+++\\
Nasal Congestion                    &+          &+          &+++\\
    \midrule
Coughing	                        &dry ++     &dry ++     &+  \\
    \midrule

Sneezing	                        &+	        &+	        &+++\\
    \midrule
Sore Throat	                        &+	        &++	        &+++\\
    \midrule
Pain: Body 	                        &+          &+++	    &++	\\
Pain: Head (Headache)	            &++ 	    &+++	    &+  \\
    \midrule
Fatigue, Tiredness	                &mild ++         &+++	    &+  \\
    \midrule

Appetite Loss                       &+          &+++        &+  \\
    \midrule
Onset Gradient                      &+          &+++        &+  \\
    \bottomrule
  \end{tabular}
\end{table}


\subsection{Monitoring of Spread}
Beyond the idea of using smartphone based surveys and AI methods to monitor the spread of the virus \cite{rao2020identification}, one could use CA for audio analysis via telephone or other spoken conversation. An AI could monitor the spoken conversations and screen for speech under cold or other symptoms as shown in Table \ref{tab:histrogram}. Together with GPS coordinates from smart phones or knowledge of the origin of the call from the cell, one could establish real-time spread maps. 

\subsection{Monitoring of Social Distancing and Effects}
Social distancing and -- in already diagnosed cases of COVID-19 or direct contact isolation of individuals -- might lead to different negative side effects. People who have less social connection might suffer from even a weaker immune system, responding less well to pathogens \cite{cole2015loneliness}. Especially, the high-risk target group of elderly could even encounter suicidal thoughts and develop depression or other clinical conditions in isolation \cite{luo2012loneliness}.
CA might provide indications about social interaction, exemplary speaking time during the day via phone or other devices, as well as measure emotions of the patient throughout the day or detecting symptoms of depression and suicidal risk \cite{cummins2015review}.

In addition, the public obedience and discipline in social-distancing could be monitored with the aid of CA. AI allows to count speakers, locate them and their potential symptoms as reflecting in the audio signal (cf.\ Table \ref{tab:histrogram}), and `diarise' the audio sources, i.\,e., attribute which symptoms came from which (human) individual. Likewise, public spaces could be empowered by AI that detects potentially risky settings, which are overcrowded, under-spaced in terms of distance between individuals, and spot potentially COVID-19 affected subjects among a crowd, and whether these and others are wearing a protective mask while speaking.    

\subsection{Monitoring of Treatment and Recovery}
During hospitalisation or other forms of treatment and recovery, CA can monitor the progress, e.\,g., by updating histograms of symptoms. In addition, the wellbeing of patients could be monitored similarly to the case of individual monitoring in social distancing situations as described above. This could include listening to their emotions, eating habits, fatigue, or pain, etc.

\subsection{Generation of Speech and Sound}
While we have focused entirely on the analysis of audio up to this point, it remains to state that there may be also use-cases for the generation of audio by AI in a COVID-19 scenario. Speech conversion and synthesis could, e.\.g., help those suffering from COVID-19 symptoms to ease their conversation with others. In such a setting, an AI algorithm can fill in the gaps arising from coughing sounds, enhance a voice suffering from pain or fatigue and further more, e.\,g., by generative adversarial networks \cite{pascual2017segan}. In addition, alarm signals could be rendered which are mnemonic and re-recognisable, but adapt to the ambient sound to be particularly audible. 

\section{Challenges}
\label{Challenges}

\subsection{Time}
The fight against COVID-19 has been marked by a race to prevent too rapid spread that could lead to peak infection rates that overburden the national health systems and availability of beds in the intensive care units leading to high morality rates. 
Further, at presence, it cannot be clearly stated whether or not COVID-19 will persistently stay as disease. However, recent research and findings \cite{wu2020characteristics} as well as model calculations indicate that COVID-19 will heavily spread over the next 6-9 months in different areas of the world. Enhanced social distancing might delay the spreading. Additionally, at the moment there is no solid research available to prove persistent immunity against the virus after an infection with COVID-19. Therefore, the need to apply measures of enhanced risk assessment, diagnosis, monitoring, and treatment is urgently necessary to support the current medical system as well as to get COVID-19 under control.

\subsection{Collecting COVID-19 Patient Data}
Machine learning essentially needs data to learn from. Accordingly, for any kind of CA application targeting speech or sounds from patients suffering from COVID-19 infection, we will need collected and annotated data. At present, such data is not publicly available for research purposes, but urgently needed. Hence, a crucial step in the first place will be to collect audio data from diagnosed patients and ideally control subjects under equal conditions and demographic characteristics.

\subsection{Model Sharing}
In order to accelerate the adaptation of machine learning models of CA for COVID-19, exchange of data will be crucial. As such data is usually highly private and sensitive in nature, the recent advances in federated machine learning \cite{yang2019federated} can benefit the exchange of personal model parameters rather than audio to everyone's benefit. Likewise, user of according services can `donate their data' in a safe and private manner.

\subsection{Real-world Audio Processing}
Most of the tasks and use-cases listed above require processing of audio under more or lass constrained `in-the-wild' conditions such as audio recording over telephone, VoIP, or audio takes at home, in public spaces, or in hospitals. These are usually marked by noise, reverberation, transmissions with potential  loss, and further disturbances. In addition, given the pandemic character of the SARS-CoV-2 Corona crisis, one will ideally need to be robust against multilingual, multicultural, or local speech and sound variability. 

\subsection{Green Processing}
Green processing summarises here the idea of efficiency in computing. This will be a crucial factor for mobile applications, but also for big(ger) data speech analysis \cite{verma2016big} such as in the case of telephone audio data analysis. It includes conservative consumption of energy such as on mobile devices, efficient transmission of data such as in the above named federated machine learning in order not to burden network transmission, memory efficiency, model update efficiency, and many further factors.

\subsection{Trustability of Results}
Machine Learning and Pattern Recognition methods as used in CA are usually statistical learning paradigms and hence prone to error. The probability of error needs to be a) estimated, known as confidence measure estimation, and b) communicated to users of CA services in the COVID-19 context to assure trustability of these methods. One step further, results should ideally be explainable. However, eXplainable AI (XAI) itself is at this time a young discipline, but provides an increasing method inventory allowing for interpretation of results \cite{adadi2018peeking}.

\subsection{Ethics}
Many of the above suggested use-cases come at massive ethical responsibility and burden which can often only be justified in times of global crisis as the current one. This includes fore mostly many of the above sketched applications of CA for monitoring. Assuring privacy at all times will be crucial to benefit only the goal of fighting COVID-19 spread without opening doors for massive miss-use. Further concerns in this context will touch upon legal and societal implications. All of these cannot be discussed here -- rather, we can provide pointers for the interested reader as starting points \cite{Batliner14-MTF,greene2019better,nebeker2019building}. 

\section{Concluding Remarks}
\label{Conclusion}
In this short overview, we provided pointers towards what Computer Audition (CA) could potentially contribute to the ongoing fight against the world-wide spread of the SARS-CoV-2 virus known as `Corona crisis' and the COVID-19 infection triggered by it. We have summarised a number of potentially useful audio analysis tasks by means of AI that have already been demonstrated feasible. We further elaborated use-cases how these could benefit in this battle, and shown challenges arising from real-life usage. The envisioned use-cases included automated audio-based individual risk assessment, audio-only-based diagnosis by symptom frequency and symptom development histograms over time in combination with machine learning, and several contributions to monitoring of COVID-19 and its effects including spread, social distancing, and treatment and recovery besides use-cases for audio generation.  At the time of writing, it seems that what matters most is a rapid exploitation of this largely untapped potential. Obviously, in this short overview, not all possibilities could be included, and many further potential use-cases may exist. Further, the authorship is formed by experts on CA, digital health, and clinicians having worked with COVID-19 infected patients over the last months -- further insights from other disciplines will be highly valuable to add. The Corona crisis demands for common efforts on all ends -- we truly hope computer audition can add a significant share to an accelerated success of the crisis' defeat.

\section{Acknowledgements}
We express our deepest sorrow for those who left us due to COVID-19; they are not numbers, they are lives. We further express our highest gratitude and respect to the clinicians and scientists, and anyone else these days helping to fight against COVID-19, and at the same time help us maintain our daily lives.
We acknowledge funding from the EU's HORIZON 2020 Grant No.\ 115902 (RADAR CNS). This work was further partially supported by the Zhejiang Lab's International Talent Fund for Young Professionals (Project HANAMI), P.\,R.\,China, the JSPS Postdoctoral Fellowship for Research in Japan (ID No.\,P19081) from the Japan Society for the Promotion of Science (JSPS), Japan, and the Grants-in-Aid for Scientific Research (No.\,19F19081 and No.\,17H00878) from the Ministry of Education, Culture, Sports, Science and Technology (MEXT), Japan.


	
 	\eightpt
	\bibliographystyle{IEEEtran}
	\bibliography{mybib}
	
\end{document}